\documentstyle[preprint,aps]{revtex}
\tighten
\begin{document}
\draft
\title{A Comparison of Polarization Observables in Electron Scattering\\
from the Proton and Deuteron}
\author{B.D. Milbrath,$^1$\thanks{Present address: Eastern Kentucky University,
Richmond, Kentucky 40475} 
J.I. McIntyre,$^2$\thanks{Present address: Rutgers University, 
Piscataway, New Jersey 08855} C.S. Armstrong,$^2$ 
D.H. Barkhuff,$^1$\thanks{Present address: Massachusetts Institute of 
Technology, Cambridge, Massachusetts 02139} W. Bertozzi,$^3$ 
D. Dale,$^3$\thanks{Present
address: University of Kentucky, Lexington, Kentucky 40506}
G. Dodson,$^3$\\ K.A. Dow,$^3$ M.B. Epstein,$^4$ M. Farkhondeh,$^3$ 
J.M. Finn,$^2$ S. Gilad,$^3$ M.K. Jones,$^2$  
K. Joo,$^3$\thanks{Present address: Thomas Jefferson National Accelerator
Facility, Newport News, Virginia 23606} J.J. Kelly,$^5$\\ S. Kowalski,$^3$ 
R.W. Lourie,$^1$\thanks{Present address:  State University of 
New York at Stony Brook, Stony Brook, New York 11794} R. Madey,$^6$ 
D.J. Margaziotis,$^4$ P. Markowitz,$^5$\thanks{Present address: Florida 
International University, Miami, Florida 33199} C. Mertz,$^7$ J. Mitchell,$^8$\\
C.F. Perdrisat,$^2$ V. Punjabi,$^9$ L. Qin,$^{10}$ P.M. Rutt,$^{11}$ 
A.J. Sarty,$^3$\thanks{Present address: Florida 
State University, Tallahassee, Florida, 32306} D. Tieger,$^3$ 
C. Tschal\ae r,$^3$\\ W. Turchinetz,$^3$ P.E. Ulmer,$^{10}$ 
S.P. Van Verst,$^{1,3}$\thanks{Present address: Washington
Dept. of Health, Division of Radiation Protection, Olympia, Washington 98504}
G.A. Warren,$^3$\thanks{Present address: Universit\"at 
Basel, CH-4056 Basel, Switzerland} L.B. Weinstein,$^{10}$ 
R.J. Woo$^2$\thanks{Present address: University of Manitoba, Winnipeg, Canada 
R3T 2N2}\\(The Bates FPP Collaboration)
}
\address{$^1$University of Virginia, Charlottesville, Virginia 22901\\
	 $^2$College of William \& Mary, Williamsburg, Virginia 23185\\
	 $^3$Massachusetts Institute of Technology and Bates Linear Accelerator
Center, Cambridge, Massachusetts 02139\\
	 $^4$California State University - Los Angeles, Los Angeles, 
California 90032\\
	 $^5$University of Maryland, College Park, Maryland 20742\\
	 $^6$Kent State University, Kent, Ohio 44242\\
	 $^7$Arizona State University, Tempe, Arizona 85287\\
	 $^8$Thomas Jefferson National Accelerator Facility, Newport News, 
Virginia 23606\\
	 $^9$Norfolk State University, Norfolk, Virginia 23504\\
	 $^{10}$Old Dominion University, Norfolk, Virginia 23529\\
	 $^{11}$Rutgers University, Piscataway, New Jersey 08855}
\date{\today}
\maketitle
\begin{abstract}
Recoil proton polarization observables were measured for both
the p($\vec {\rm e}$,e$^\prime\vec{\rm p}\,$) 
and d($\vec {\rm e}$,e$^\prime\vec{\rm p}\,)$n reactions 
at two values of Q$^2$ using a newly commissioned proton Focal Plane 
Polarimeter at the M.I.T.-Bates Linear Accelerator Center.
The hydrogen and deuterium spin-dependent observables $D_{\ell\ell}$ and 
$D_{{\ell}t}$, the induced polarization $P_n$ and the form factor
ratio $G^p_E/G^p_M$ were measured under identical kinematics. The deuterium 
and hydrogen results are in good agreement with each other and with the 
plane-wave impulse approximation (PWIA).
\end{abstract}

\narrowtext

For many years, a major effort of nuclear physics has 
been the determination of the nucleon electromagnetic 
form factors.  In the Breit frame, the Sachs representation of the elastic 
form factors, $G_E$ and $G_M$, represent Fourier transforms of the charge and 
magnetization densities of the nucleon; the same interpretation is also 
obtained at low Q$^2$ in the nucleon rest frame.
Precise experimental determination of these form factors imposes
stringent constraints on models of baryon structure.

In the past, the Q$^2$-dependence of the proton form 
factors\cite{AndBos,Hohler,Walker,Bartel,Janssens,ArnSil} has 
been measured using the Rosenbluth separation technique.
Extracting the form factors requires performing a set of measurements at 
fixed Q$^2$ while varying the electron scattering angle $\theta_e$ and the 
incident electron beam energy $E$.
Because the technique relies on absolute cross-section measurements, it is 
sensitive to systematic errors in $E, E^\prime$ (the scattered
electron energy), and $\theta_e$.

The Rosenbluth separation technique has also been used to determine the
Q$^2$-dependence of the neutron form factors via quasielastic electron-deuteron 
scattering\cite{Hughes,GroJul,BraHas,Platchkov}.  Consequently, the extraction 
of the neutron information is sensitive to deuteron wave function models.  
It appears possible, however, using polarization 
techniques (such as d($\vec {\rm e}$,e$^\prime\vec{\rm n}\,)$p\cite{Eden})
to determine the neutron form factors in a nearly model-independent fashion.
This requires that polarization observables measured on the 
deuteron for quasifree kinematics be insensitive to specifically nuclear 
mechanisms such as Final State 
Interactions (FSI), Meson Exchange Currents (MEC), and Isobar Configurations 
(IC). Since recoil polarimetry can be used for both neutrons and
protons, it is possible to test these assumptions using the complementary 
reaction d($\vec {\rm e}$,e$^\prime\vec{\rm p}\,)$n and directly compare the 
results to those obtained using recoil polarization in elastic proton 
scattering.

In the p($\vec {\rm e}$,e$^\prime\vec{\rm p}\,$) reaction, there are, 
assuming one-photon exchange, two helicity-dependent polarization 
observables\cite{Arnold,Akhiezer,Dombey}:
\begin{eqnarray}
P_t=&hD_{{\ell}t}=&{h\over I_0}\left(-2\sqrt{\tau(1+ \tau)}\,
G^p_{M} G^p_{E}\tan{\theta_e\over 2}\right)\nonumber\\
P_\ell=&hD_{\ell\ell}=&{h(E + E^\prime)\over I_0 M_p}\sqrt{\tau(1+\tau)}\,
(G^{p}_{M})^{2}{\tan}^2{\theta_e\over 2}\quad.\label{pols}
\end{eqnarray}
The subscripts $t$ and $\ell$ refer to the recoil proton's polarization 
components in the electron scattering plane, either transverse or longitudinal 
to its momentum.  The first subscript in the polarization transfer 
coefficients $D_{{\ell}t}$ and $D_{{\ell\ell}}$ refers to the electron's 
longitudinal polarization.  The electron beam helicity is denoted by $h$, 
$I_0$ is the unpolarized cross section (excluding $\sigma_{Mott}$), and $\tau =
Q^2/4M$.

The measurement of polarization observables is of interest because they
result from the interference between amplitudes and thus may be linear in 
small, interesting quantities rather than quadratic (as in cross-section 
measurements).  An example of this is $G_E^p$ in $P_t$.  It is increasingly 
difficult to measure $G^p_E$ as Q$^2$ becomes large 
(\hbox{$>$\kern-0.8em\lower.95ex\hbox{$\sim$}} 1 (GeV/c)$^2$) 
using Rosenbluth separation because the cross section is kinematically
dominated by $G^p_M$. 
Note that the ratio of $P_t/P_\ell$ gives the ratio of the
form factors $G^p_E/G^p_M$ independent of the beam helicity:
\begin{equation}
{G^p_E \over G^p_M} = -{P_t \over P_\ell}{(E + E^\prime)
\tan{\theta_e\over 2}\over 2M_p}\quad.\label{ratio}\end{equation}
Because these two polarization observables are measured simultaneously, this 
technique avoids a major systematic uncertainty of the Rosenbluth method.

A recoil polarization component normal to the (e,e$^\prime$) plane,
$P_n$, may, for example, be induced by FSI.  Such a
polarization is helicity independent, unlike the above longitudinal and
transverse polarizations.  For elastic scattering from a proton, $P_n$
vanishes in one-photon exchange.  Comparing the measured polarization 
observables in both p($\vec {\rm e}$,e$^\prime\vec{\rm p}\,$) and 
d($\vec {\rm e}$,e$^\prime\vec{\rm p}\,)$n scattering allows a sensitive, 
model-independent test of the impulse approximation for the deuteron.

The experiment\cite{Bates8821,McIntyre,Milbrath} was performed at the 
M.I.T.-Bates Linear Accelerator Center during the winter of 1995.  
A longitudinally polarized electron beam of 580 MeV with a current ranging 
from 5-15 $\mu$A and a 1\% duty factor was incident on a cryogenic target. 
The target had cells for both liquid 
hydrogen and deuterium. The hydrogen and deuterium 
target cells were 5 and 3 cm in diameter, respectively. They were
alternated in the beam every 8--12 hours.  The scattered 
electrons were detected in the Medium Energy Pion Spectrometer (MEPS) 
while the scattered protons were detected in the One-Hundred Inch Proton 
Spectrometer (OHIPS).  Both spectrometers contain two focussing quadrupoles 
followed by a vertically-bending dipole.  MEPS had a 14 msr solid angle 
acceptance while OHIPS had a 7.0 msr solid angle acceptance.  The momentum 
acceptances were
$\pm 10\%$ and $\pm 5\%$, respectively.  A focal plane polarimeter (FPP) 
built by the experimenters was installed on OHIPS, allowing the polarization 
of the outgoing protons to be measured.
Data were acquired at two different 
electron scattering angles, 82.7$^\circ$ and 113$^\circ$ corresponding
to four-momentum transfers squared of 0.38 and 0.50 (GeV/c)$^2$.

The FPP consists of four two-plane multiwire proportional chambers, two each 
before and after a graphite analyzer, allowing the proton trajectory to be 
determined both before and after it scatters in the graphite.  The analyzer 
thickness (7 cm and 9.5 cm for the 0.38 and 0.50 (GeV/c)$^2$ Q$^2$ 
measurements, respectively) was chosen to optimize the figure of merit.  
Scattering angles $\theta$ in the graphite could be 
resolved to $\le1^\circ$ and the FPP provided complete azimuthal coverage for
$\theta\le20^\circ$. The device was calibrated in a direct beam of
polarized protons at the Indiana 
University Cyclotron Facility (IUCF) in February of 1993\cite{LourieIUCF}.

The angular distribution of the $^{12}$C(p,p$^\prime$) scattering in the
analyzer, in terms of focal plane polarizations, is\cite{Aprile}
\begin{equation}
I(\theta,\phi)=I_0(\theta)\big[1-P_n^{fp}A_c(\theta)\sin\phi+P_t^{fp}A_c
(\theta)\cos\phi\big]\quad,\label{secscat}
\end{equation}
where $I_0(\theta)$ is the unpolarized angular distribution, $\phi$ is the
second scattering azimuthal angle and 
$A_c$ is the analyzing power of the $^{12}$C(p,p$^\prime$) reaction.
$A_c$ depends on the second
scattering polar angle $\theta$ and the proton kinetic energy $T_p$.  
$A_c$ peaks between 10-20$^\circ$ and goes to zero as $\theta$ goes to 
zero because the small angle scatterings are predominantly spin-independent 
multiple Coulomb scatterings; however, the $^{12}$C(p,p$^\prime$) elastic 
cross-section is dominated by these small-angle 
(\hbox{$<$\kern-0.8em\lower.95ex\hbox{$\sim$}} 3.5$^\circ$) events.  
To eliminate such events the read-out electronics was equipped with a fast 
small-angle rejection system described elsewhere\cite{LourieNIM}.

The small-angle rejection system implemented a box cut on the second 
scattering coordinates $x$ and $y$, resulting in azimuthally-biased small 
angle data. This bias was removed by a software cut that excluded events 
scattering through less than 7$^\circ$.  Events with 
$\theta >20^\circ$ were also excluded since $A_c$ is not well known at larger 
angles.  Instrumental asymmetries of the FPP were separated from the physical
asymmetries by elastically scattering unpolarized electrons from
hydrogen. Any $P_n$ 
component to this data could only result from two or more photon
exchange and would thus be negligible in comparison to instrumental
effects; therefore, we treated any such component as an
instrumental asymmetry and subtracted it 
from the $P_n$ component of the deuterium data.
                        
The $A_c$ values used to extract the physical asymmetries were 0.514 and 0.537 
for the 0.38 and 0.50 (GeV/c)$^2$ Q$^2$ measurements, respectively.  These were
determined 
using a fit of the form developed by Aprile-Giboni {\it et al.}\cite{Aprile} 
on a database that included our IUCF calibration measurement and other similar
measurements\cite{Aprile,McNaughton}.  The uncertainty in the measured proton
polarization due to the analyzing power was 1.4\% for the lower Q$^2$ 
measurement and 1.9\% for the higher Q$^2$ measurement.

The electron beam polarization was measured on a daily basis using a M\o ller
polarimeter.  It could also be determined from the hydrogen data using the 
$G^p_E/G^p_M$ ratio as determined from the FPP.  This ratio was used in 
eq. \ref{pols} to determine a value of $D_{{\ell}t}$.  By then taking the 
ratio $P_t/D_{{\ell}t}$ the helicity was determined.
These results agreed with the M\o ller data to within 2.0\% and are shown in 
table \ref{table1}.  The first error bars are statistical
while the second are systematic.  

The FPP measures only the two polarization components
perpendicular to the proton momentum vector; however, they are each determined 
for both helicity states ($+$ and $-$) so that there are four observables at 
the focal plane:  two helicity sums and two helicity differences 
[$(P^+_{fp} \pm P^-_{fp})_{i=1,2}$].  Because $P_{\ell}$ 
and $P_t$ are helicity-dependent while $P_n$ is not, all three polarization
components at the target can then be extracted by exploiting the spin mixing 
in the spectrometer magnets.

In order to extract the polarization components at the target from the focal 
plane polarizations it was necessary to model the spin precession in OHIPS.
This was done utilizing the optics code COSY\cite{COSY} which
generated a trajectory- and energy-dependent spin precession matrix
${\bf M}$ such that ${\bf P}_{fp} = {\bf MP}_{tgt}$.  To the extent that 
${\overline {\bf M\cdot P_{tgt}}} \approx {\bf {\overline M}\cdot P_{tgt}}$ 
(found to differ by less than 1\% in a Monte Carlo simulation using a realistic
model of the deuteron), the method of least 
squares can be used to give the maximum likelihood estimate of the three 
polarization components at the target in terms of the four focal plane 
observables\cite{Barkhuff}.  In a separate analysis,
the Monte-Carlo program MCEEP\cite{MCEEP} was coupled
with several physics models\cite{Arenhovel,PicVan} to generate
polarized scattering events appropriately weighted by their production 
cross-section over the full experimental acceptance.  Using ${\bf M}$, the 
polarization vector for each of these 
events was transported to the focal plane and their ensemble average then 
compared to the experimental data. Recoil polarizations at the target 
extracted using these two different methods were consistent with one another 
to better than $0.6$\%.

To facilitate the comparison between the hydrogen and deuterium data,
the recoil momentum of the residual neutron for the deuterium data was 
restricted to the range 0-60 MeV/c.  
A precise subtraction of the polarization of accidental events
was made for the deuterium data.

Table \ref{table2} summarizes the experimental results for the hydrogen
and deuterium targets.  
The first error bars are 
statistical while the second are uncorrelated systematic errors due to 
kinematic uncertainties and also uncertainties in the positions of the 
spectrometer magnets which affect the spin precession.  
Figure \ref{figure1} compares these results with previous Rosenbluth separation 
measurements.  The error bars represent the statistical and systematic errors added 
in quadrature.  Our deuterium (solid 
diamonds) results are slightly offset in Q$^2$ from our hydrogen (solid circles)
measurements to allow comparison.  The data are in good agreement with 
previous Rosenbluth measurements.  The previous measurements shown in the 
figure are from H\"ohler {\it et al.}\cite{Hohler} (open circles), 
Bartel {\it et al.}\cite{Bartel} (open square), and 
Janssens {\it et al.}\cite{Janssens} (Xs).  The dot-dash\cite{Mergell} 
and short-dashed\cite{Simon} curves are based on vector dominance models  
while the long-dashed curve\cite{GarKru} is based on an extended vector 
dominance 
model.  Table \ref{table3} shows the measured polarization observables 
$D_{\ell\ell}$, $D_{{\ell}t}$ and $P_n$ for the proton and deuteron.  Their
systematic errors include, in addition to the previously mentioned kinematic and
magnet position uncertainties, larger correlated errors due to
uncertainties in the beam polarization (4\%, which does not affect $P_n$) and the
analyzing power.

The hydrogen and deuterium data agree with each other, which precisely 
confirms the validity of the Impulse Approximation at these kinematics.
The deuteron data are consistent with theoretical calculations by Arenh\"ovel
assuming a dipole parameterization of the form factors 
that predicts negligible influence from FSI, MEC, and 
IC at our kinematics.\cite{Arenhovel}
We have demonstrated that recoil polarization observables may be
precisely determined at intermediate energies and, as
these observables are inherently much more sensitive than
spin-averaged ones to the presence of small amplitudes, this technique 
shows great promise for future measurements of, for example,
$G_E^n$\cite{Madey} and $G_E^p$ at higher Q$^2$\cite{PerPun}.

We would like to thank the staff at the Bates Linear Accelerator Center for 
their assistance in carrying out this experiment, as well as 
H. Arenh\"ovel for his calculations of the deuteron polarization observables.  
One of us (B.D.M.) would like to acknowledge the support of the Air Force 
Office of Sponsored Research.  This work was supported in part by the 
Department of Energy under Grants Nos. DE-FG05-90ER40570 and DE-FG05-89ER40525,
and by the National Science Foundation under Grants Nos. PHY-89-13959, 
PHY-91-12816, PHY-93-11119, PHY-94-11620, PHY-94-09265, and PHY-94-05315.  
One of us (R.W.L.) acknowledges the support of a NSF Young Investigator Award.

%
%
\begin{figure}
\caption{The ratio $\mu_p G_E/G_M$ for both the proton (solid circles) and 
deuteron (solid diamonds) vs. Q$^2$.  The error bars represent the 
statistical and systematical errors added in 
quadrature.  The fits and other (Rosenbluth) data are listed in the text.  
The deuterium data are offset slightly for the sake of clarity.}
\label{figure1}
\end{figure}

%
%
\narrowtext
\begin{table}
\caption{Summary of beam helicity measurements.}
\label{table1}
\begin{tabular}{lcc}
Device&$h$ (Q$^2 = 0.38$ (GeV/c)$^2$)&$h$ (Q$^2 = 0.50$ (GeV/c)$^2$)\\
\tableline
FPP&$0.281\pm0.014\pm0.004$&$0.275\pm0.013\pm0.006$\\
M\o ller&$0.287\pm0.002\pm0.012$&$0.280\pm0.002\pm0.011$\\
\end{tabular}
\end{table}

\narrowtext
\begin{table}
\caption{Summary of $\mu_p G^p_E/G^p_M$ measurements.}
\label{table2}
\begin{tabular}{lcc}
Reaction&(Q$^2 = 0.38$ (GeV/c)$^2$)&(Q$^2 = 0.50$ (GeV/c)$^2$)\\
\tableline
p($\vec {\rm e}$,e$^\prime\vec{\rm p}\,$) & $1.016\pm0.052\pm0.016$& 
$0.970\pm0.047\pm0.020$\\
d($\vec {\rm e}$,e$^\prime\vec{\rm p}\,)$n& $1.024\pm0.103\pm0.016$& 
$1.005\pm0.064\pm0.021$\\
\end{tabular}
\end{table}

\widetext
\begin{table}
\caption{Summary of polarization transfer coefficients.  The theoretical
calculations are by Arenh\"ovel.}
\label{table3}
\begin{tabular}{lclll}
Reaction&Q$^2$ ((GeV/c)$^2$))&$D_{\ell\ell}$&$D_{{\ell}t}$&$P_n$\\
\tableline
p($\vec {\rm e}$,e$^\prime\vec{\rm p}\,$)&0.38&$0.627\pm0.031\pm0.027$& 
$-0.510\pm0.007\pm0.022$&$0.0000\pm0.0022\pm0.0000$\\
d($\vec {\rm e}$,e$^\prime\vec{\rm p}\,)$&0.38&$0.624\pm0.060\pm0.027$& 
$-0.513\pm0.016\pm0.022$&$-0.0014\pm0.0042\pm0.0000$\\
d($\vec {\rm e}$,e$^\prime\vec{\rm p}\,)_{theory}$&0.38&0.649&$-$0.508&
$-$0.0033\\
p($\vec {\rm e}$,e$^\prime\vec{\rm p}\,$)&0.50&$0.858\pm0.030\pm0.038$& 
$-0.410\pm0.014\pm0.019$&$0.0002\pm0.0042\pm0.0000$\\
d($\vec {\rm e}$,e$^\prime\vec{\rm p}\,)$&0.50&$0.825\pm0.038\pm0.037$& 
$-0.408\pm0.018\pm0.019$&$-0.0045\pm0.0052\pm0.0001$\\
d($\vec {\rm e}$,e$^\prime\vec{\rm p}\,)_{theory}$&0.50&0.866&$-$0.422& 
$-$0.0024\\
\end{tabular}
\end{table}

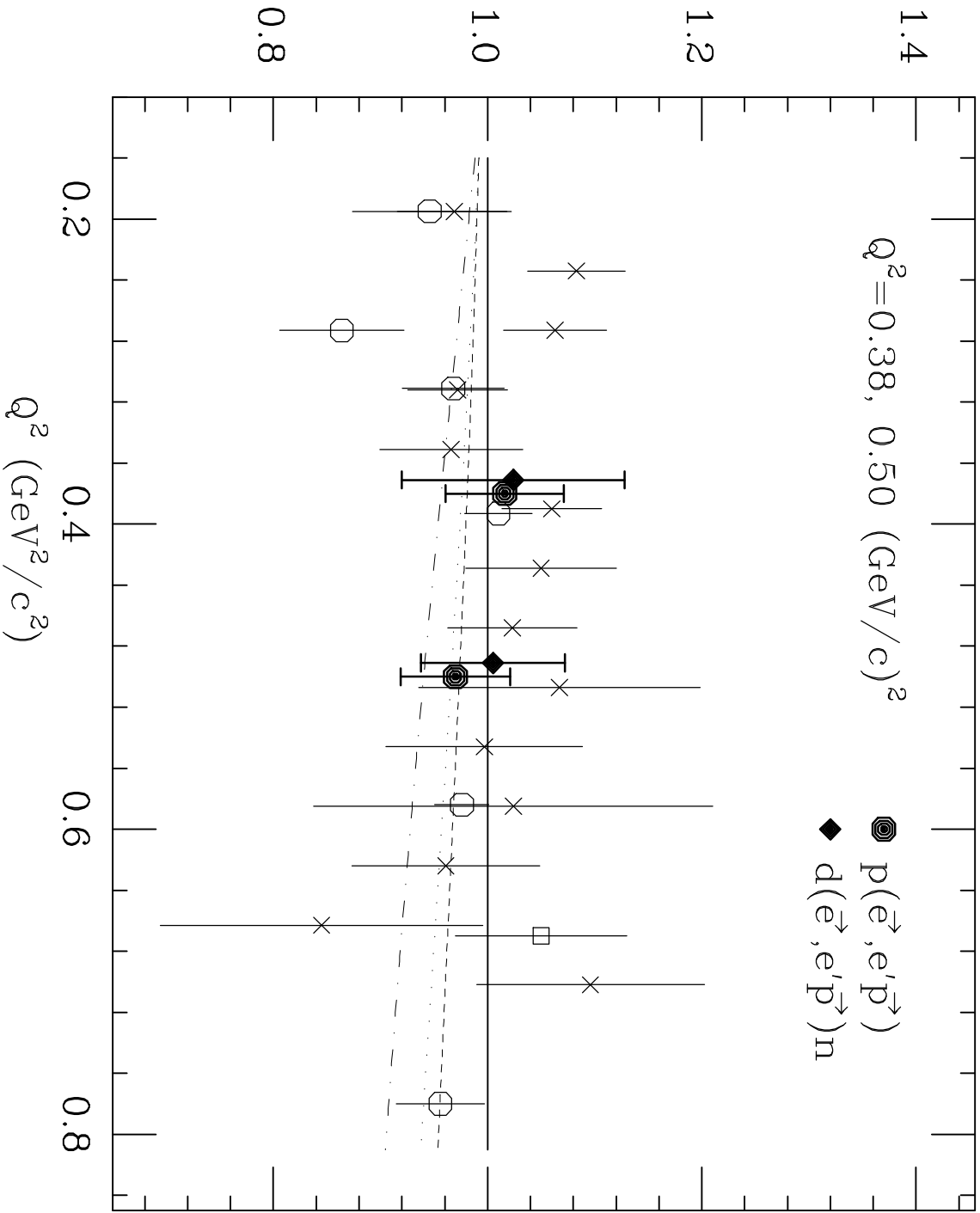
\end{document}